\documentclass[10pt]{iopart}

\usepackage{iopams}
\usepackage{graphicx}
\begin{document}
\title[]{Characterization of disturbance sources for LISA: torsion pendulum results}\author{L Carbone$^a$, A Cavalleri$^b$, R Dolesi$^a$,
C D Hoyle$^{a}$\footnote[1]{Presently at University of Washington, Seattle (WA), USA},\\ M Hueller$^a$, S Vitale$^a$ and W J Weber$^a$}

\address{$^a$ Dipartimento di Fisica, Universit\`a di Trento and INFN, Sezione di Padova, Gruppo Collegato
di Trento, I-38050, Povo, Trento, Italy }
\address{$^b$ CEFSA-ITC, I-38050, Povo, Trento, Italy}

\ead{carbone@science.unitn.it}

\begin{abstract}
A torsion pendulum allows ground-based investigation
of the purity of free-fall for the LISA test masses inside their
capacitive position sensor.
This paper presents recent improvements in our torsion pendulum facility
that have both increased the pendulum sensitivity and allowed detailed characterization
of several important sources of acceleration noise
for the LISA test masses. We discuss here an improved upper limit 
on random force noise originating in the sensor. Additionally, we present new
measurement techniques and preliminary results for characterizing the forces caused by the sensor's
residual electrostatic fields, dielectric losses, residual spring-like coupling, and temperature gradients.  
\end{abstract}
\section{Introduction}
Achieving the LISA gravitational wave sensitivity requires that the test masses (TM) are kept 
in free fall with residual accelerations below 3 fm s$^{-2}$ Hz$^{-1/2}$,
in the low frequency band from 0.1Hz down to 0.1mHz \cite{bender:LISA}.
This is achieved by a Drag-Free control scheme, where external disturbances are shielded by
the spacecraft, which is driven to be centered
about the freely-falling TM according to a Gravitational Reference Sensor (GRS).
In a high feedback loop 
regime with gain $\omega_{DF}^2$, the TM acceleration noise $a_n$ can be divided in two
main categories, as shown as follows:
\begin{equation}
a_{n}=\frac{f_{str}}{m}+\omega^2_p\left({x_{n}+\frac{F_{str}}{M\omega_{DF}^2}}\right)
{\label{a:noise}}
\end{equation}
Here, position independent stray forces $f_{str}$
directly act on the TM (mass $m$), whereas spring-like couplings $k_p=m \omega^2_p$ (``stiffness'')
between the TM and the spacecraft (mass M) couples the residual relative motion,
due to position sensor noise $x_n$ or external forces $F_{str}$ acting on the spacecraft, into acceleration noise.

Gravitational Reference Sensors, based on a capacitive readout and actuation scheme, have been
developed to meet LISA requirements in terms of both position and acceleration noise 
\cite{weber:sensor,dolesi:sensor}. In the ground-testing programme, proceeding in parallel
to the design of the LISA flight test LTP \cite{borto:LTP},
a torsion pendulum facility has been developed and is used
to validate the functionality of capacitive position sensor prototypes and, 
most importantly, to investigate 
the purity of free fall allowed by the GRS \cite{hueller:pend,carbone:prl,hueller:upperlimits}.
Moreover, with its high force sensitivity this facility is also used to accurately characterize several possible
disturbance sources that could prevent the achievement of the LISA goal.

Recent studies \cite{carbone:prl,hueller:upperlimits} have already
put significant upper limits on surface stray forces,
including those related to the read-out back-action electrostatic noise and molecular impacts.
Additionally, the most significant source of spring-like coupling, 
the AC voltage bias used for the capacitive position readout, has been measured
and found to agree with the theoretical model prediction at the 10\% level. 
The need for a more detailed characterization of the GRS 
focuses current experimental activity on improving 
the torsion pendulum noise and extending its sensitivity to
lower frequencies. 
Moreover, it investigates several noise sources which have been identified
as threatening to LISA. 

In this paper we report on most recent results of this experimental campaign.
In the first section we will focus on the upgrades of the facility.
Then we will discuss the most recent upper limits 
on stray forces, allowed by the pendulum's improved sensitivity.
Finally,
we will describe measurement techniques developed to 
investigate stray electrostatics effects in the GRS, like electrical patch fields or dielectric losses,
thermal gradients related effects and residual spring-like couplings.
 
\section{Apparatus Design}
A detailed description of the entire facility is given in 
\cite{hueller:upperlimits}.
The torsion pendulum is comprised of a representative copy of a LISA TM 
hanging from a long thin fiber 
inside a Mo-Shapal GRS prototype 
\cite{weber:sensor,dolesi:sensor},
all inside a high vacuum chamber. 
We use a hollow cubic TM, made of Au-coated Ti, with 40 mm side length and 2 mm wall thickness,
and a bare W fibre, 1 m long and 25 $\mu$m thick.
This pendulum configuration
is rather insensitive to net forces and to magnetic or gravitational bulk  effects,
but it maximizes the sensitivity to surface effects, likely the most dangerous for the sensor under developement.
The pendulum inertia is I = 3.48 g cm$^2$ and the
torsional spring constant is $\Gamma\approx$ 5.5 nN m rad$^{-1}$ 
with a mechanical quality factor Q $\approx$ 3000. 
The resulting resonance frequency
is f$_0\approx$ 2 mHz.

Any torque $N(\omega)$ acting on the TM can be
detected as deflection of the pendulum angular rotation $\phi(\omega)=F(\omega)N(\omega)$, 
through the transfer function 
$F(\omega) = \left(\Gamma \left( 1 - (\omega/\omega_0)^2 + i/Q \right) \right)^{-1}$. The measured 
torque noise $S^{1/2}_N(\omega)=\left|F(\omega)\right|^{-1}S^{1/2}_\phi$,
where $S^{1/2}_\phi$ is the pendulum angular noise, 
sets upper limits on stray forces $S^{1/2}_f(\omega)=S^{1/2}_N(\omega)/ R_\phi $, 
making use of a suitable conversion armlength $R_\phi$ depending 
on the specific class of the noise source \cite{carbone:prl,hueller:upperlimits}. 
Additionally, the high torque sensitivity allows precise 
measurements of individual noise sources
by coherent modulation of the disturbance source itself.

\begin{figure}[t]
\begin{center}
\includegraphics[width=11cm]{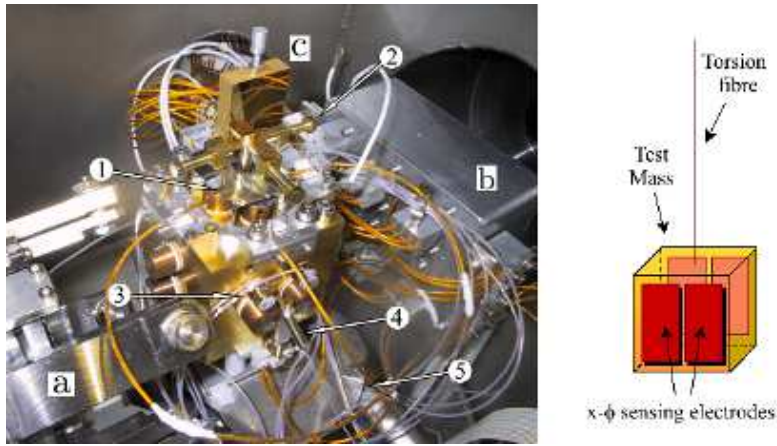}
\caption{{\label{sensor}} On the left, a photograph of the capacitive sensor inside the vacuum chamber,
integrated on the micromanipulator for the positioning (a) and wired 
with the read-out electronics box (b). The mirror for the independent optical
read-out is visible (c). Most recent upgrades are labeled:
Electrical shields (1); Tunable test mass support (2); Electrical heaters (3);
UV light optical fiber (4); Motorized rotary stage (5).
On the right, a sketch of the suspended TM surrounded by the $x$ and $\phi$ 
sensing electrodes.}
\end{center}
\end{figure}

In preparation for the current experimental run, several hardware upgrades
were implemented, as shown in \fref{sensor}, to improve the pendulum noise performances
and allow a detailed investigation of individual force noise sources.
The entire pendulum suspension, which is electrically isolated from the TM, has been
Au-coated for electrostatic homogeneity. Additionally, tuning screws have been added to minimize
the mass quadrupole moment of the rectangular mirror used for the optical readout, reducing possible coupling to
gravitational disturbances. Replacement of a stainless steel capillary tube used
for holding the pendulum with a Cu one reduced the pendulum's residual
magnetic moment to 7 nA m$^2$. 
Electrical shields cover dielectric
surfaces of area $\approx$ mm$^2$ facing the TM support.
This shields undesirable stray couplings
which have produced a ``trans-twist'' effect \cite{carbone:prl,hueller:upperlimits},
converting sensor-TM relative translation displacement into a torque.
The ``trans-twist'' couplings, $\frac{\partial N}{\partial \theta}$ 
and $\frac{\partial N}{\partial \eta}$,
have been reduced 
by more than one order of magnitude down to several nN,
and thus no longer limit the pendulum noise performance.

The sensor prototype sits on a motorized rotational stage, allowing
coherent modulation of the sensor-TM relative rotation angle,
needed for the characterization of the overall spring-like couplings.
A set of four electrical heaters and five precision thermometers is attached
on electrode housing, to induce thermal gradients in order to 
investigate the related effects. Finally, a
simplified discharge system, 
composed by UV lamps, optical fibers
and vacuum feedthroughs has been provided by Imperial College of London,
and was integrated in the torsion pendulum facility to test the functionality 
of the charge control scheme under development for LISA \cite{sumner:charge}.

\section{{ Characterization of force noise sources}}
{\bf Upper limits on stray forces.} 
\begin{figure}[t]
\begin{center}
\includegraphics[width=100mm]{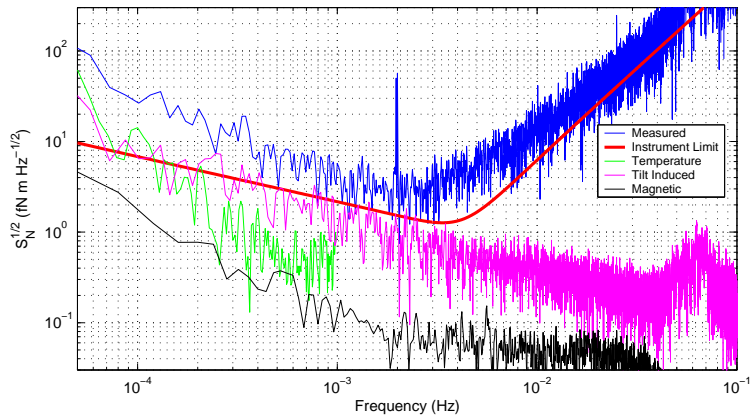}
\caption{\label{torque:noise} A typical torque noise spectrum of the torsion pendulum (blue line) 
for a time range of 140000s.
For comparison we show the instrument limit (red),
given by pendulum's intrinsic thermal noise in low frequency region, 
and by the readout noise at higher frequencies. 
The peak at 2mHz is artificially produced by the 69000s Hamming window 
applied to calculate the spectrum.
The torque noise induced by temperature fluctuations 
in the column surrounding the 
torsion fiber is shown in green.
The pink line shows the torque noise 
induced by the residual ``trans-twist'' coupling described in \cite{carbone:prl,hueller:upperlimits},
while magnetic torque noise is shown in black.
}
\end{center}
\end{figure}
The typical torque sensitivity of the torsion pendulum is plotted in \fref{torque:noise}. 
For comparison, we also show the instrument limit 
$S^{1/2}_N(\omega)=\sqrt{ S_{N_{th}}(\omega) + S_{\phi_{read}} \left| F(\omega) \right|^{-2}} $,
where $S^{1/2}_{N_{th}}(\omega)=\sqrt{4k_BT \Gamma / (\omega Q)}$ is the pendulum's
intrinsic thermal noise, and $S^{1/2}_{\phi_{read}}\left| F(\omega) \right|^{-1}$ 
is the capacitive sensor angular read-out noise
converted into equivalent torque noise.
At frequencies above 0.4mHz, measured noise is below 
10 fN m Hz$^{-1/2}$, 
with a minimum noise floor of 3 fN m Hz$^{-1/2}$ around 2mHz. 
The source of the observed excess noise,
between a factor 1.5 and 4 from the instrument limit
in this frequency range, is currently under investigation.
These data allow us to set significant upper limits
on the acceleration noise for LISA.
Following \cite{hueller:upperlimits}, 
we convert the torque noise into acceleration noise
as if our hollow 40 mm TM were a solid 1.3 kg Au/Pt of the same dimensions. 
Above 0.4 mHz, for a class of noise like circuitry back action,
for which R$_\phi$ = 10.25 mm (the nominal half separation between adjacent electrodes),
acceleration noise is below 1 pm s$^{-2}$ Hz$^{-1/2}$,
with a minimum about 250 fm s$^{-2}$ Hz$^{-1/2}$ around 2 mHz.
For homogeneously distributed noise sources
including random inelastic molecular impacts, where R$_\phi$ = 20 mm,
this minimum is $\approx$ 115 fm s$^{-2}$ Hz$^{-1/2}$,
about a factor 40 from the LISA goal.
At 0.1mHz, the lowest frequency of the LISA band, 
we estimate a force noise upper limit of order of 1.5 pN Hz$^{-1/2}$ for this class of sources.
\\
{\bf Stray DC biases.}
Stray DC electrostatic fields, due to different work functions 
of the sensor conducting surfaces, surface contamination 
or imperfectly shielded external fields,
interact with the fluctuating TM charge
to produce acceleration noise.
Methods to measure
and compensate stray DC bias imbalances down to the sub-mV level during the flight missions
have been proposed \cite{weber:dc} and experimentally verified\footnote[2]{For LISA,
the relevant DC bias imbalance is the average translational potential difference
across the sensor $\Delta_x$, whereas with the pendulum
we measure the analogous rotational imbalance $\Delta_\phi$ \cite{weber:dc}.}
in lab \cite{carbone:prl}.  
The measurement technique simulates a sinusoidally varying TM charge by applying 
dither voltages to selected electrodes and produces on the TM a force (and torque) 
proportional to the same DC bias imbalances that can produce acceleration noise.
The average biases, and thus the resulting force,
can be nulled by purposely applying DC compensation voltages to the sensing electrodes through
the actuation circuitry. 
This allows a reduction of DC biases of a factor 
$\approx$ 100 compared to 
measured uncompensated values,
that we found to be typically of order tens of mV across the sensor housing,
and yields a suppression of
the related acceleration noise down to a negligible level for LISA.
\begin{figure}[t]
\begin{center}
\includegraphics[width=100mm]{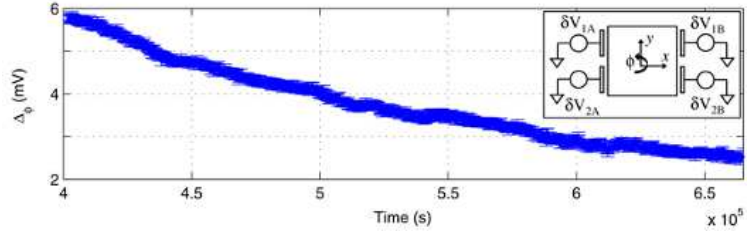}
\caption{\label{dcbias:time} Time series of a 3 days measurement of
one of the stray DC bias combinations, namely 
$\Delta_\phi=\left(\delta V_{1a} + \delta V_{2b} - \delta V_{2a} - \delta V_{1b}\right)$ as
shown in the inset and defined in \cite{weber:dc}.
A compensation of the -62.8 mV average imbalance is applied.  }
\end{center}
\end{figure}
\begin{figure}[t]
\begin{center}
\includegraphics[width=100mm]{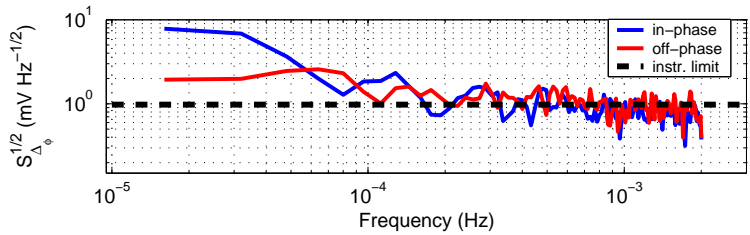}
\caption{\label{dcbias:noise} Noise spectrum of the DC bias time series showed
in \fref{dcbias:time}. 1000 data points, corresponding to 250000s, are analyzed using
a 62500s Hamming window.
As described in the text, 
the in-phase demodulated signal (blue line), which carries the $\Delta_\phi$ information,
and off-phase (red line) signal, which should contain only noise, show up with the same noise
levels down to 0.1 mHz. The instrument limit is shown for comparison (black dashed line),
and is given by the pendulum torque noise at the measurement frequency of 4mHz.}
\end{center}
\end{figure}

During this experimental campaign, we performed a complete investigation 
of all DC bias imbalances 
characteristic of the Mo-Shapal sensor prototype under testing.
In addition to the net DC bias imbalances, performance limiting effects 
can also arise in the DC bias fluctuations.
As described in \cite{stebbins:errors} for the case of fluctuating actuation voltages,
a fluctuation in the stray voltage imbalance $\Delta_x$,
along the LISA sensitive axis,
mixes with the TM net charge $q$ to  
produce force noise on the TM
$S^{1/2}_F\approx \frac{\partial C}{\partial x} \frac{q}{C_T} S^{1/2}_{\Delta_x}$,
with $\frac{\partial C}{\partial x}$ and $C_T$ defined in \cite{weber:dc}.
Additionally, unstable elements of $\Delta_x$ mix 
with the individual biases themselves
to produce force noise. 
The LISA requirement for fluctuating
voltages $\approx$10 $\mu$V Hz$^{-1/2}$,
which produces a maximum acceleration noise $\approx$ 0.2 fm s$^{-2}$ Hz$^{-1/2}$
with both charge and net DC bias contributions taken into account \cite{stebbins:errors},
can not be measured yet at this level with this pendulum.
An example of measurement of the long term stability of 
the DC bias imbalance $\Delta_\phi$, which is the rotational analogous
of $\Delta_x$, is plotted in \fref{dcbias:time}.
The related noise spectrum is shown in \fref{dcbias:noise}, where the 
in-phase demodulated signal, which carries the $\Delta_\phi$ information, and 
the off-phase one, which should contain only noise, are analyzed. 
An upper limit can be set at 1mV Hz$^{-1/2}$, a factor 100 over the LISA requirement. 
For frequencies higher than 0.1 mHz, the measured
noise is limited
by the pendulum torque noise at the measurement frequency of 4mHz.
The excess voltage noise at lower frequency
is indicative of an observed non-stationary drifting of the DC bias,
as illustrated in \fref{dcbias:time}.
This excess noise does not depend on the applied compensation voltages, nor is
related with possible fluctuations of the dithering voltages used to perform the measurement. 
To investigate a possible correlation, DC bias measuremenst have been
performed at different temperatures and appling different temperature gradients.
No clear sistematic dependence have been observed in both cases,
but changing temperature seems to destabilize the net imbalance $\Delta_{\phi}$. 
Currently, an aggressive investigation is ongoing 
to better understand the origin of these stray DC voltages and their fluctuations, 
as well as their spatial distribution in the sensor.
\\
{\bf Test Mass Charge Control.}
In LISA, the TMs become electrically charged by cosmic ray particle radiation, 
undergoing Coulomb forces for the interaction with surrounding and nominally grounded surfaces
and Lorentz forces due to its motion in the interplanetary magnetic field.
The method to actively control the charge uses the emission of photoelectrons
by UV light in combination with suitable electron driving voltages \cite{sumner:charge}. 
A charge measurement technique, where dither voltages are applied to the sensing electrodes to
produce a coherent torque proportional to the TM voltage drop $V_M$ to
the grounded sensor housing \cite{weber:dc}, has been successfully tested 
with the torsion pendulum facility. Using a typical modulation frequency f$_m$ = 5 mHz,
and a dither bias amplitude of 50 mV, a 3 hour measurement allows to resolve 
$V_M$ at the sub-mV level, corresponding to $\approx$ 5000 elementary charges.
This resolution is well below the level
of 10$^6$ elementary charges envisioned
as a discharge threshold, before the charge-induced stiffness,
to be discussed in the next section,
becomes dominant \cite{bender:LISA}.
Furthermore, a preliminary test of this discharging scheme has been performed 
using the UV light hardware provided by the Imperial College of London.
The UV light fiber is aligned to illuminate both the suspended TM and one of the sensing electrodes.
The two photoelectron currents emitted from both surfaces can be driven
from the TM to the electrode or viceversa
by purposely applying bias voltages to the electrode plates.
Preliminary tests, performed with a crude fibre alignment
and with a simple on/off control of the UV light intensity,
demonstrated the capability of bipolar charge transport and control.
The observed charge transfer rate
is at least of order of 10$^5$ elementary charges per second.
The lower limit is imposed by the slow sampling time 
($\approx$ 200 s) of the charge measurement and
by the slow torsion pendulum dynamics.\\
{\bf Stiffness.}
As shown in \eref{a:noise}, spacecraft-TM springlike couplings (``stiffness'') $k_p$ converts
any residual relative spacecraft motion
into force noise on the TM. 
As discussed in \cite{weber:sensor,dolesi:sensor}, electrostatics play a leading role
among the possible stiffness sources, and the 
likely dominant contribution is given by the 
100 kHz sensing bias of the capacitive readout.
The rotational analogous of the sensing stiffness has been already measured and found 
to agree with theoretical predictions at the 10\% level \cite{carbone:prl,shaul:sensor}, 
nevertheless it is also important to search for any unmodeled coupling 
arising in the sensor.
\begin{figure}[t]
\begin{center}
\includegraphics[width=100mm]{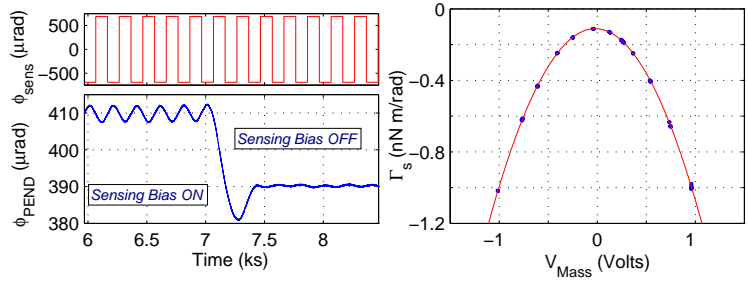}
\caption{\label{stiffness} 
Left Panels: A sequence of two measurements of the total rotational stiffness,
respectively with
the sensing bias on and off. On top, the sensor rotation angle
about the suspended test mass, as read by the motorized stage encoder,
modulated at 5 mHz. 
On the bottom, the pendulum 
motion, where the free pendulum oscillation
and higher harmonics of the modulation frequency are filtered.
The pendulum coherent response, as well as the equilibrium position, 
change when removing the sensing bias stiffness.
Right panel: Measurement of the rotational stiffness as function of the TM voltage drop to the sensor housing (blue dots). The red line
shows the quadratic fit. Uncertainties for each stiffness data point are about $\approx$1 pN m rad$^{-1}$.}
\end{center}
\end{figure} 
 
An example of the measurement technique developed to investigate
the total rotational stiffness $\Gamma_s$ induced by the GRS is given in
\fref{stiffness}. Square wave 
modulation of the sensor angle $\phi_{sens}$, by means of the 
rotary stage, induces a torque 
$N(t)=-\Gamma_s\left(\phi - \phi_{sens}(t)\right)$
on the pendulum.
The torque amplitude, and thus the stiffness itself,
is measured by the coherent deflections of the pendulum angular rotation.

As shown in \fref{stiffness}, the total stiffness has been measured both with the sensing 
bias on and off. 
By the difference of the two measurements, the sensing bias stiffness
contribution results to be (-89.2 $\pm$ 0.4) pN m rad$^{-1}$ 
in agreement with previous results \cite{carbone:prl}.
Besides, a residual stiffness $\Gamma_{res}=$ (-11.6 $\pm$ 0.2) pN m rad$^{-1}$ has been measured.
This unpredicted coupling could be explained by patch voltages of order of 100 mV$_{rms}$ homogenously
distributed across the sensor housing. Note that this rms value is consistent with
the average DC bias
typically measured.
Another possible explanation is a residual of the ``trans-twist'' interaction,
through a coherent sensor translational displacement of order of $\mu$m associated with mrad angular motion, due to mounting misalignment. 
A further analysis is ongoing to investigate the source of this residual stiffness.

A potentially relevant contribution to the stiffness arises from Coulomb forces
between the electrically charged TM and the surrounding 
sensor housing conduction surfaces. As for the sensing bias,
the measurement of the charge dependent stiffness is a key verification
of the sensor electrostatic model. 
Measurements described in the previous paragraph
were performed holding $V_M$=0 V.
For this investigation, 
the TM has been biased to different voltage drops $V_M=q/C_T$
exploiting its finite
discharge time ($\approx$ 10 hours) through the rest 
of the pendulum, by directly biasing the
W torsion fiber\footnote[6]{This finite discharge time would require continuous charge control
in order to hold $V_M$ fixed. To avoid the electrostatic stiffness associated with 
electron driving voltage affecting the measurements,
the UV light charge control scheme is not employed in this experiment.}. 
Results of this investigation are shown in \fref{stiffness}.
The measured coupling
$\Gamma_s (V_M)=-\frac{\partial N}{\partial \phi}=-\frac{1}{2} \sum_i \left(\frac{\partial^2 C_i}{\partial \phi^2}\right)V_M^2$,
where $ \sum_i \left(\frac{\partial^2 C_i}{\partial \phi^2}\right)$ = (1.84$\pm$ 0.01) nF rad$^{-2}$ 
is the sum of the second derivatives of all electrode capacitances $C_i$ with respect to the TM,  
is in agreement within 10\% with the F.E.M. prediction \cite{shaul:sensor}. It is worth 
noting that this estimate is not completely independent from the model prediction,
because of the assumption of $\frac{\partial C_x}{\partial \phi}\simeq$11.5pF rad$^{-1}$,
estimated with F.E.M analysis, 
which enters in the calibration of the charge measurement.

The maximum stiffness\footnote[7]{As $\Gamma_s$ is negative, 
the maximum value corresponds to the minimum coupling.} 
is found for $V_M$=(-20.2 $\pm$ 0.3) mV, 
rather than $V_M$=0 V as one might expect. 
This is likely due to the difference between the "zero"
defined by average potential of the four electrodes
relevant to the charge measurement and the weighted
average of the potential on all conducting surfaces
for which $\frac{\partial^2 C_i}{\partial \phi^2}\not=0$ (relevant to the stiffness measurement).
This 20 mV difference is consistent with measured levels of DC biases in the sensor.

Beyond the successfull confirmation of the reliability of the electrostatic model,
it is worth nothing that a $\approx$ 100 mV$_{rms}$ stray DC voltage, a possible cause of
the measured residual stiffness, would produce a translational stiffness
in LISA of order of $\approx$10 nN m$^{-1}$. This is about 10 times smaller
than the dominant contributions envisioned, and thus it would not
create a significant noise source for the scientific mission. 
\\
{\bf Dielectric Losses.}
\begin{figure}[t]
\begin{center}
\includegraphics[width=100mm]{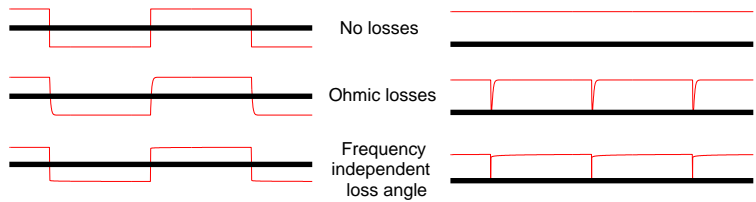}
\caption{{\label{losses}} Correspondence between voltage applied $V(t)$ (left) and torques $N(t)\propto V^2(t)$ induced on the 
pendulum (right) in the case of zero loss, ohmic loss and frequency independent loss angle.}
\end{center}
\end{figure}
Voltage thermal noise 
originates in dielectric lossy layers on electrodes and TM surfaces, characterized by a dielectric loss angle $\delta$,
and couples to DC voltages to produce force noise on the TM itself.
In order to keep this contribution whithin 10\% of the LISA
acceleration noise budget,
the maximum allowed is $\delta \approx 10^{-5}$\cite{stebbins:errors}.

Up to now, our attempts to measure $\delta$
through its effect on the pendulum Q, as in \cite{speake:damping},
have been unreproducibile 
at a level which prevents the extraction $\delta$ to better than 10$^{-4}$.
To overcome this limit, we 
developed a new measurement technique,
which takes advantage of the transient produced 
on square wave voltages applied on a lossy conductor surface.
The principle of this technique is schematized in \fref{losses}.
In the lossless ideal case,  
a square wave voltage,
applied to the sensing electrodes, would induce a constant torque 
$N(t)\propto V^2(t)$ on the TM.
In the presence of losses, 
torque transients are produced coherently with each voltage transition.
The torque results in a train of pulses, whose amplitude is
strictly related to the loss itself, at twice the voltage carrier frequency.

A first calibration of the method has been performed 
by adding a resistor to create a true ohmic loss, characterized 
by a frequency independent time delay $\tau$.
For the case of a frequency independent loss angle
(for which $\tau\propto\omega^{-1}$),
a synthetized time series simulating the waveform for a fixed $\delta$
has been used to calibrate. 
In both cases, resulting measured torques are in agreement with expected 
values with accuracy below the 10\% level.
\begin{figure}[t]
\begin{center}
\includegraphics[width=100mm]{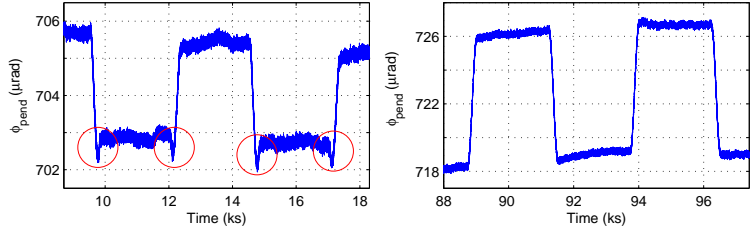}
\caption{{\label{losses:meas}} Examples of the dielectric loss angle measurement technique. 
In both cases, the free pendulum mode signal is filtered and the 0.2mHz signal is due to uncompensated DC bias.  
The left panel shows a calibration performed
using ohmic losses (corresponding to $\tau\approx 10 ms$): 
the pendulum response at the torque
transient produced at 0.4mHz is highlighted with red circles. 
The right panel shows the pendulum response during a sensor related $\delta$ measurement:
the small signal, measured by coherent demodulation at 0.4 mHz and corresponding to $\delta\approx 10^{-6}$, is
here not readily visible.}
\end{center}
\end{figure}
With the sensitivity shown in \fref{torque:noise}, the method 
has a resolution of the order of $10^{-16}$ N m for a few hour measurement
and permits extraction of $\delta$ to within 10$^{-6}$. 

Measurements have been performed
on different electrode pairs, for voltage amplitudes
from 1 V to 3 V, and 
at different frequencies between 0.2 and 3mHz.
Preliminary results give an estimate of $\delta\approx10^{-6}$,
over the frequency range of investigation.
A more detailed model of ohmic and dielectric losses
in the sensor is needed in analyzing theese measurements.
Still, it is worth noting that
these values are one order of magnitude
less than the maximum dielectric losses allowed for LISA.
\\
{\bf Thermal Gradient Related Effects.}
Temperature gradients $\Delta T$  across the sensor housing, with spectral density $S^{1/2}_{\Delta T}$, are expected to 
cause noisy forces onto the TM surface area A, according to the following formula:
\begin{equation}\label{t:noise}
S^{1/2}_{F_{\Delta T}}= \left( \frac{8\sigma}{3 c} A T^3 +\frac{A}{4} \frac{P}{T}+ \frac{A}{4} G \mathcal{Q} \frac{\Theta}{T^2 }\right) S^{1/2}_{\Delta T}
\end{equation}
Here, the first two terms are related, respectively, to the radiation pressure effect, where 
$\sigma$ is the Stefan-Boltzmann constant and $c$ is the speed of light, and to the radiometric effect,
linearly proportional to the average pressure $P$ inside the sensor, 
which can be greater than the pressure in the vacuum chamber due to the outgassing of the 
electrode housing internal surfaces.
The last term is due to temperature dependent outgassing,
causing pressure difference across the TM proportional to the temperature gradient $\Delta T$.
Here $\mathcal{Q}$ is the outgassing rate inside the sensor, scaled by the geometrical factor $G$ resulting 
from a combination of the conductance of the paths around the test mass and through the 
holes in the electrode housing walls, and $\Theta$ is an effective activation temperature 
of the outgassing phenomena. 
With a temperature gradient stability of $S^{1/2}_{\Delta T}\approx$ 10 $\mu$K Hz$^{-1/2}$,
the current LISA error budget tentatively
apportions an acceleration noise $\approx$ 0.5 fm s$^{-2}$ Hz$^{-1/2}$ \cite{stebbins:errors},
almost equally distributed among the different envisioned contributions.
The uncertainty on this estimate 
arises mainly on the lack of knowledge of the parameters $\Theta$ and $\mathcal{Q}$, which 
necessitates an experimental investigation, which would, moreover,
highlight any other possible
unexpected thermal related disturbance.

\begin{figure}[t]
\begin{center}
\includegraphics[width=120mm]{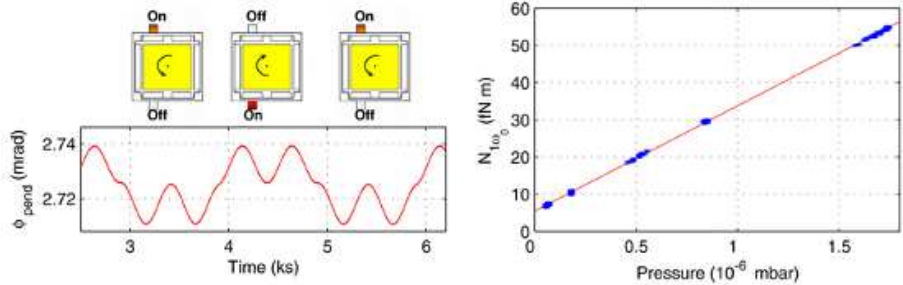}
\caption{\label{thermal} Left panels: an example of the pendulum coherent response (bottom) to the 
heat of 0.2 Watt alternative applied to opposing electrodes every 1000s (top).
We roughly estimated that the consequent temperature distribution inside the sensor is equivalent,
from the point of view of the torque produced, to the presence of 70 mK temperature drop between
the corresponding two opposite electrodes (with an uncertainty of a factor 2).
Right Panel: Demodulated torques at the carrier frequency of 0.5mHz 
measured as function of the residual pressure in the vacuum vessel.
The average sensor temperature for these measurements is T$\approx$ 295 K. }
\end{center}
\end{figure}

The contributions from each effect should be distinguishable 
by its specific dependence on pressure and temperature. 
In particular, any pressure independent signal larger 
that the radiation pressure effect would represent an 
important upper limit for the less known outgassing related effects.
A preliminary investigation has been performed 
here by applying off-axis ``rotational''
temperature differences
that convert the forces described above 
into torques measurable with this pendulum (see \fref{thermal})
The results have been compared with the prediction of 
a simplified model that allows for converting the 
forces described in \eref{t:noise} into torques. This model assumes that the  
effects are isotropic (dependent just on the local temperature)
and a rough estimation of the temperature distribution 
inside the sensor, based on the measurement performed 
by a thermometers array positioned on its external surfaces 
(its accuracy  will improve thanks to  the currently in 
progress thermal F.E.M. of the sensor).
Measurements of torque, coherently induced by alternatively 
applying 0.2 W to the heaters behind two opposing electrodes 
at a frequency of 0.5 mHz,  are shown in \fref{thermal} as function 
of the pressure measured  inside the vacuum vessel. The linear 
dependence is consistent with the presence of the radiometric 
effect, which is the leading term above $\approx 2\times 10^{-7}$ mbar.
The observed slope 
is $2.8 \times  10^{-8}$ N m mbar$^{-1}$ and is compatible  with our rough  
prediction of  $3 \times 10^{-8}$ N m mbar$^{-1}$ for radiometric effect,
affected by an inaccuracy of a factor 2. 
The pressure independent terms show up in the positive
intercept at zero pressure: being the expected radiation 
pressure responsible for half of it, the  unknown outgassing 
related effects turn out to give a comparable  contribution.
These results are in line with of the assumption of the current 
LISA noise budget, and
show that there are no contributions,
from outgassing or other effects,
that are likely to be present 
at levels well in excess of
the radiation pressure and radiometric effects. 
However, it is advisable to confirm them 
with a more detailed and accurate investigation with a 
four masses pendulum configuration \cite{hoyle:4mass}, 
which is sensitive 
directly to the net forces acting on the TM.
Such a measurement is less model dependent and guaranties more accurate 
estimation of a possibly non-uniform outgassing contribution. 

\section{Conclusions}

In the near future  
we will use the torsion pendulum facility to 
characterize a new implementation of GRS,
which is presently in preparation,
employing the current LTP geometry design for position
sensor and TM (4mm TM-sensing electrodes gaps and 46mm TM cubic size, different electrode configuration \cite{weber:sensor}).

Additionally, a four mass pendulum facility \cite{hoyle:4mass}, which is currently under development
and that will be built in few months, will allow a more representative characterization 
of the GRS related force disturbances along the translational 
axis of relevance for LISA.

\ack{The authors would like to thank T.J.Sumner, G.K.Rochester and D.N.A. Shaul
at the Imperial College of London for 
their collaboration in suppling
the charge management hardware and electrostatic modelling
of the sensor.
This work was supported by ESA, INFN and ASI.}

\end{document}